\documentclass[sigplan]{acmart}

\acmConference[arXiv]{}
\acmYear{}
\acmISBN{} 
\acmDOI{} 
\startPage{1}
\setcopyright{none}

\usepackage{alltt}
\usepackage{xspace}

\newcommand*{\yaps}{\textsc{Yaps}\xspace}
\newcommand*{\ycode}[1]{\mbox{\small\texttt{#1}}}
\newcommand*{\samp}{\ycode{\textbf{\textasciitilde}}\xspace}
\newcommand*{\lsamp}{\ycode{\textbf{<}}\samp}

\begin{document}

\settopmatter{printacmref=false} 
\renewcommand\footnotetextcopyrightpermission[1]{} 
\pagestyle{plain} 

\title{\yaps: Python Frontend to Stan}

\author{Guillaume Baudart \qquad Martin Hirzel \qquad Kiran Kate \qquad Louis Mandel \qquad Avraham Shinnar}
\affiliation{\institution{IBM Research, USA}}
\email{guillaume.baudart@ibm.com,hirzel@us.ibm.com,kakate@us.ibm.com,lmandel@us.ibm.com,shinnar@us.ibm.com}

\begin{abstract}
  Stan is a popular probabilistic programming language with a
self-contained syntax and semantics that is close to graphical models.
Unfortunately, existing embeddings of Stan in Python use multi-line
strings. That approach forces users to switch between two different
language styles, with no support for syntax highlighting or simple
error reporting within the Stan code. This paper tackles the question of
whether Stan could use Python syntax while retaining its
self-contained semantics. The answer is yes, that can be accomplished
by reinterpreting the Python syntax. This paper introduces \yaps, a
new frontend to Stan based on reinterpreted Python. We tested \yaps on
over a thousand Stan models and made it available open-source.

\end{abstract}

\maketitle
\renewcommand{\shortauthors}{Baudart, Shinnar, Hirzel, Mandel, Kate}

\section{Introduction}\label{sec:intro}

A \emph{probabilistic model} is a mathematical model for explaining
real-world observations as being generated from latent
distributions~\cite{ghahramani_2015}.  Probabilistic models can be
used for machine learning, and compared to alternative approaches,
have the potential to make uncertainty more overt, require less
labeled training data, generate synthetic data, and improve
interpretability~\cite{baudart_hirzel_mandel_2018}.  The key
\emph{abstractions} for writing probabilistic models are
\emph{sampling} of latent variables and observations and
\emph{inference} of latent variables~\cite{gordon_et_al_2014}.
Probabilistic modeling is supported by several stand-alone
domain-specific programming languages, e.g.,
Stan~\cite{carpenter_et_al_2017}.  On the other hand, machine learning
is supported by many Python-based packages.  To capitalize on Python's
packages and familiarity, PyStan~\cite{carpenter_et_al_2017},
PyCmdStan~\cite{woodman_2017}, and other
efforts~\cite{uber_2017,salvatier_wiecki_fonnesbeck_2015,tran_et_al_2017}
embed probabilistic abstractions into Python.

In programming, a \emph{watertight abstraction} provides a basis for
coding or debugging without \emph{leaking} information about
lower-level abstractions that it builds upon. Unfortunately,
probabilistic abstractions in Python offered by existing efforts~\cite{uber_2017,salvatier_wiecki_fonnesbeck_2015,tran_et_al_2017} are
not watertight~\cite{baudart_hirzel_mandel_2018}. To code with those
packages, one must also use lower-level packages such as NumPy,
PyTorch, or TensorFlow. Furthermore, bugs such as tensor dimension
mismatches often manifest at those lower levels and cannot be reasoned
about at the probabilistic level alone.
This paper introduces \yaps (Yet Another Pythonic Stan), a watertight
embedding of Stan into Python. \yaps is available as open-source code
(\url{https://github.com/ibm/yaps}).

\begin{figure}
  \hspace*{-0.025\columnwidth}\includegraphics[width=1.05\columnwidth]{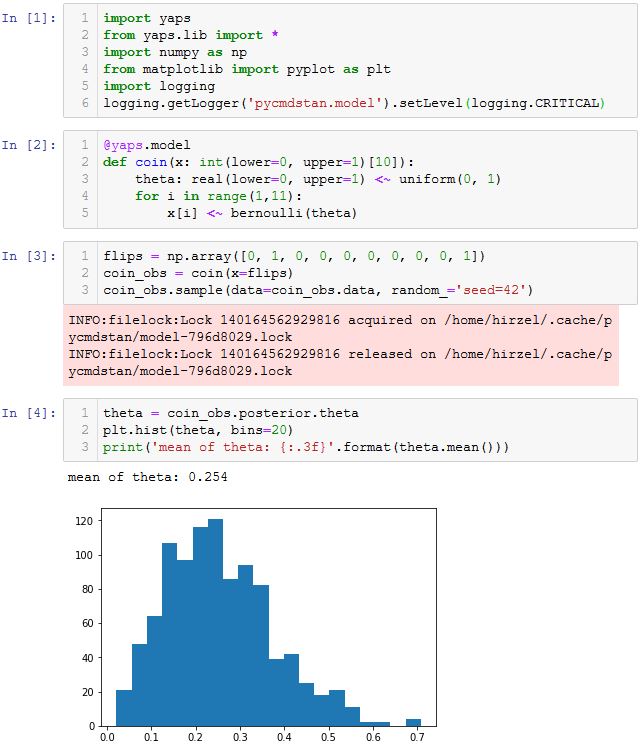}~\\
  \hspace*{-0.025\columnwidth}\includegraphics[width=1.05\columnwidth]{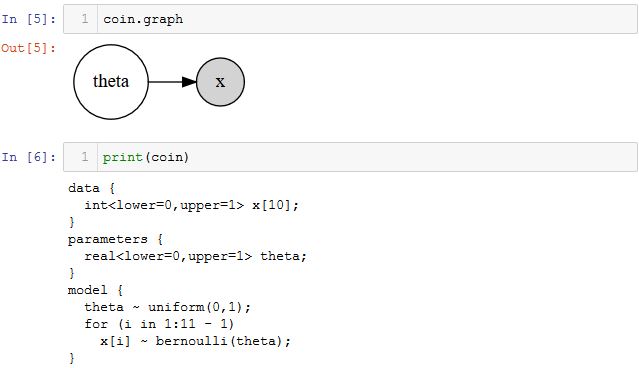}
  \caption{\label{fig:coin_yaps}Coin example in \yaps.}
\end{figure}

Figure~\ref{fig:coin_yaps} shows a \yaps example in a Jupyter
notebook~\cite{perez_2014}. Cell~1 imports \yaps and other Python
packages and makes PyCmdStan less verbose.
\mbox{Cell 2 Line 1} uses the \ycode{@yaps.model} decorator to indicate
that the following function, while being syntactically Python, should
be semantically reinterpreted as Stan. Since the code is
reinterpreted, its original Python interpretation is no longer
available, and thus, does not leak abstractions. Line~2:2 declares
\ycode{coin} as a probabilistic model with one observed
variable~\ycode{x}, representing ten coin tosses, each of which is
either tails~(0) or heads~(1). The type \ycode{int(lower=0,
    upper=1)[10]} comes from Stan, where it is used for informative
error messages and compiler optimizations. Line~2:3 declares a latent
variable \ycode{theta} for the unknown bias of the coin. The
initialization \ycode{\lsamp uniform(0, 1)} samples
\ycode{theta} with the prior belief that any bias is equally
likely. \mbox{Lines 2:4--2:5} indicate that each of the coin tosses
\ycode{x[i]} is sampled from a Bernoulli distribution with the
same latent \ycode{theta} parameter.

While watertight abstractions are essential for modeling, interaction
with the host language and other libraries is essential for
probabilistic inference and for exploring the results of inference.
\mbox{Cell 3 Line 1} represents concrete observed coin flips using the
popular NumPy package~\cite{vanderwalt_colbert_varoquaux_2011}.
Line~3:2 passes the observed data as an
actual argument to the model. And Line~3:3 calls inference to infer a
joint posterior distribution using the \ycode{sample} function of
PyCmdStan, with a random seed for reproducibility~\cite{woodman_2017}.
Cell~4 shows how to interactively explore the results of
inference. \yaps creates a \ycode{posterior} object with fields for
latent variables such as \ycode{theta}.  \mbox{Cell 4 Line 1}
retrieves the posterior distribution of \ycode{theta}.  Line~4:2 uses
the popular MatPlotLib package~\cite{hunter_2007} to plot its
histogram, and Line~4:3 prints its mean. The inferred posterior belief
is that the coin is biased towards tails, with a mean \ycode{theta}
of~0.254. That makes sense, since the observed data from \mbox{Cell 3
  Line 1} contained more tails than heads.

To reinterpret Python syntax, we must first parse it, and once parsed,
we can easily visualize its dependencies. Cell~5 yields a visual
rendering of the graphical model. \yaps compiles to Stan code, and Cell~6
prints the result, showing Stan's explicit code blocks
for observations (\ycode{data}), latent variables
(\ycode{parameters}), and the actual model. In Python, we opted
for a more concise (but equally watertight) syntax with implicit
blocks inspired by SlicStan~\cite{gorinova_gordon_sutton_2018}.

This paper argues that even when a domain-specific language such as Stan is
embedded into a host language such as Python, it should avoid leaky
abstractions and gratuitous use of host-language strings.  This paper
shows how to accomplish these objectives via reinterpreted Python,
while also providing high-quality error messages. But
the main contribution is a practical one: this paper introduces \yaps,
a new Python-embedded frontend to Stan.

\section{Related Work}\label{sec:relatedwork}

\begin{figure}
  \hspace*{-0.025\columnwidth}\includegraphics[width=1.05\columnwidth]{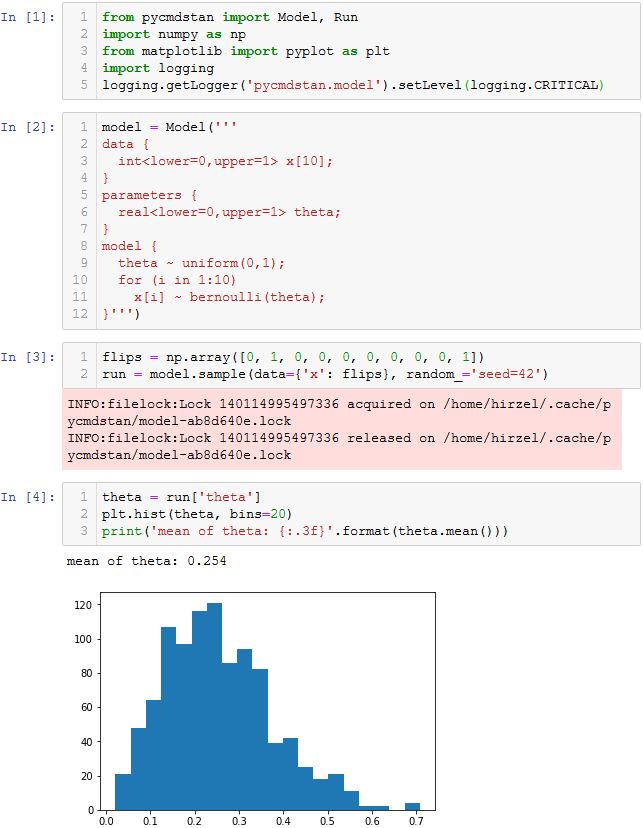}
  \caption{\label{fig:coin_pycmdstan}Prior work: coin example in PyCmdStan.}
  \vspace*{-3mm}
\end{figure}

PyStan~\cite{carpenter_et_al_2017} and PyCmdStan~\cite{woodman_2017},
the existing Python interfaces for Stan, focus on communication with
the inference engine.  They read Stan models from a file or from a
multi-line Python string and send them to the Stan compiler. Building
on this bridge, \yaps adds an additional layer, supporting Python
syntax for writing models.  Figure~\ref{fig:coin_pycmdstan} shows the
PyCmdStan version of \mbox{Cells 1--4} of the \yaps example from
Figure~\ref{fig:coin_yaps} (\mbox{Cells 5--6} illustrate \yaps
features not present in PyCmdStan).  The \yaps version is shorter than
the PyCmdStan version.  Both versions produce the same results because
the actual Stan model in \mbox{Figure \ref{fig:coin_pycmdstan} Cell 2}
is equivalent to the model generated by the \yaps compiler in
\mbox{Figure \ref{fig:coin_yaps} Cell 6}.

There are various other Python-embedded probabilistic programming
languages, such as PyMC3~\cite{salvatier_wiecki_fonnesbeck_2015},
Edward~\cite{tran_et_al_2017}, and Pyro~\cite{uber_2017}. Compared
to these, \yaps is more
watertight. Those efforts use \emph{lazy evaluation}: they overload
operators that appear to do eager computation to instead generate a
computational graph to be evaluated during inference. Lazy evaluation
is a popular approach for embedding domain-specific languages into
host
languages~\cite{hudak_1998,rompf_odersky_2012}. Unfortunately, lazy
evaluation in Python does not track local variable names, does not
cleanly isolate the embedded language, and leads to verbose syntax
that is less similar to stand-alone probabilistic languages. In
contrast, the reinterpretation approach of \yaps avoids those
disadvantages.  Both approaches (lazy evaluation and reinterpretation)
have the advantage of working in pure Python without a separate
preprocessor.

Like our work, several other recent efforts also recognize the
potential to reinterpret Python code by parsing it.  Both Tangent and
Myia build a computational graph from which they derive derivatives
that are crucial for gradient-descent based machine
learning~\cite{wiltschko_vanmerrienboer_moldovan_2017,breuleux_vanmerrienboer_2017}.
And Relay builds a computational graph both for automatic
differentiation and for mapping to heterogeneous
hardware~\cite{roesch_et_al_2018}. In contrast to these papers, our
paper focuses on a watertight embedding of a probabilistic programming
language in Python.

\section{Design and Implementation}\label{sec:design}

\paragraph{Language design and rationale.}
The design of \yaps follows the motto ``Stan-like for
probabilistic features, Python-like for everything else''.  \yaps uses
familiar Python syntax for non-probabilistic features such as
\ycode{for} loops, type declarations, or function declarations.
Stan-specific features are expressed with syntactically-valid Python syntax
that resembles the original syntax: \lsamp for sampling
or $x\!$\ycode{.T[a,b]} for truncated distribution.
\yaps uses the syntax \lsamp for sampling because it
resembles \samp from stand-alone probabilistic languages such as Stan,
is concise, and is syntactically valid
Python (infix operator \ycode{\textbf{<}} followed by prefix
operator~\samp).

Since observed variables are free during modeling but bound during
inference, \yaps make them formal arguments of the model
(Figure~\ref{fig:coin_yaps} Cell~2 Line~2) and actual arguments when
preparing the model for the inference call (Figure~\ref{fig:coin_yaps} Cell~3 Line~2). Here,
reinterpretation enabled more intuitive syntax.
Whereas Stan
requires users to place functions, parameters, transformed parameters, model,
transformed data, and generated quantities into separate code blocks,
\yaps uses SlicStan-style program
analysis~\cite{gorinova_gordon_sutton_2018} to let users place
these at the top-level in the function.  This syntax is more concise
and flexible and SlicStan shows it can even improve modularity.  For
users who prefer explicit blocks, \yaps offers a syntax based on
Python \ycode{with} statements.

\begin{figure}[!h]
  \includegraphics[width=0.95\columnwidth]{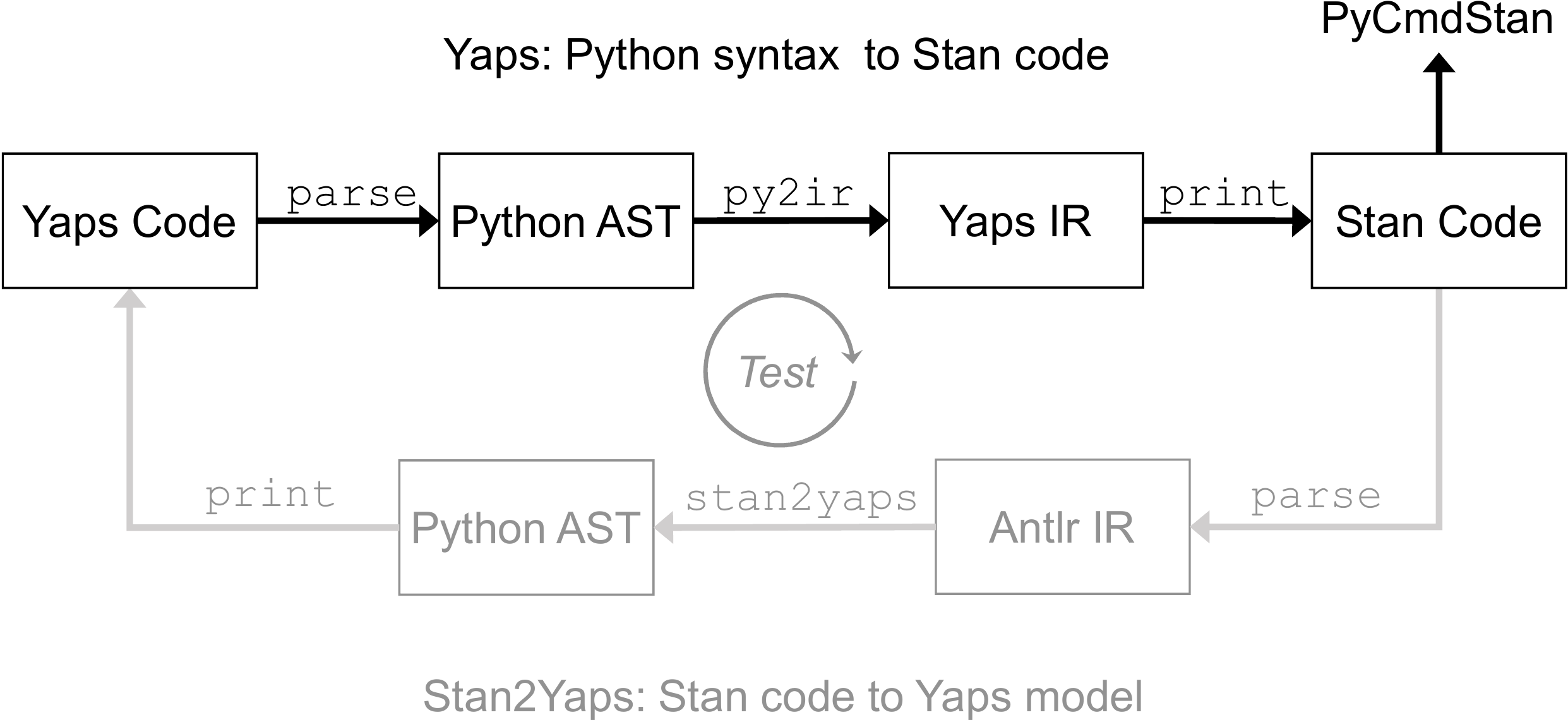}
  \caption{\label{fig:compiler}Compiler stages and representations.}
\end{figure}

\paragraph{Implementation difficulties and solutions.}
The top portion of Figure~\ref{fig:compiler} shows the stages and
representations of the \yaps compiler.
Writing our own embedded Python parser would have been cumbersome, but
fortunately, the Python standard library modules \ycode{inspect}
and \ycode{ast} solved that for us.  Our \ycode{@yaps.model}
decorator uses those modules to replace the Python function by an
intermediate representation (IR) suitable for
visualization, compilation to Stan, and inference.  One difficulty was to
implement a suitable syntax for sampling.  We first tried
\ycode{\textbf{=}}\samp, where \ycode{\textbf{=}} is Python's assignment, but that was
insufficient, because the left-hand side of Stan's sampling is more
expressive than that of Python's assignment~\cite{burroni_et_al_2018}.
Hence we settled on
\lsamp instead, which we substitute with \ycode{is} before parsing,
since that does not occur in Stan and has a low precedence in Python.

Another difficulty was name resolution.
Identifiers in \yaps code refer to Stan types, functions, and
distributions; for watertight abstractions, they should not be
resolved to Python entities.  The Python interpreter does not attempt
to resolve names in \yaps function bodies. The Python interpreter
does report errors for unknown names in function signatures, e.g.,
\ycode{def model2(N: int, y: real[N])}.  To avoid that, \yaps provides
stubs (e.g., \ycode{int}, \ycode{real}) and enables users to declare
other required identifiers (e.g.
\ycode{N = yaps.dependent\_type\_var()}).

Some tokens that are identifiers in Stan are keywords in Python, such
as \ycode{lambda}. \yaps users cannot use them as identifiers in their
models.

Finally, building a robust interface between Python and the Stan
compiler would also have been cumbersome, but fortunately,
PyCmdStan~\cite{woodman_2017}
solved that for us. The only difficulty was that PyCmdStan error messages
refer to locations in generated Stan code.  We implemented a reverse
source location mapping and used that to make error messages refer to
source locations in \yaps code instead.

\section{Evaluation}\label{sec:evaluation}

We conducted three kinds of tests to evaluate \yaps. First, we tested
our compiler on a large number of programs; second, we performed runtime tests on \yaps code; and third, we tested
whether \yaps yields good error messages.

To obtain many realistic programs to test our compiler on, we built a
second compiler that goes from Stan to \yaps, shown at the bottom of
Figure~\ref{fig:compiler}. This \ycode{stan2yaps} compiler can be used
to import existing Stan models to \yaps to ease the transition. In
this paper, we used the round-trip of the two compilers to enable the
following experiments:
\begin{itemize}
  \item We extracted 61 examples from the official Stan manual. The
    round-trip test succeeded for all of them~(100\%).
  \item We tried the round-trip test on 721 programs from the Stan dev
    repository. It succeeded for 700 of them ($97\%$). The failed
    tests used deprecated syntax.
  \item We tried the round-trip test on 500 programs from the Stan
    examples repository. It succeeded for 411 of them ($82\%$). Again,
    the failed tests used deprecated syntax.
\end{itemize}

To test whether the output of \yaps code matches the output of equivalent 
Stan code, we performed runtime tests as follows: we picked 13 Stan models with corresponding datasets and 
used our round-trip setup as above to generate Stan code through \yaps. 
We then compared the output of the \ycode{sample} function of
PyCmdStan~\cite{woodman_2017} on the original Stan model code and the generated Stan model code. 
The output matched for all of the 13 models.

\begin{figure}
  \includegraphics[width=0.95\columnwidth]{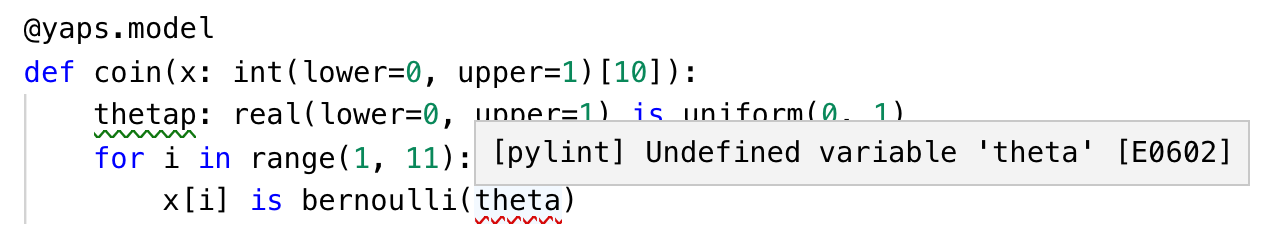}
  \caption{\label{fig:error_tools}Error reporting with existing Python tools.}
\end{figure}

To test whether \yaps improves error messages, we conducted two
experiments. First, we intentionally misspelled a variable name and
looked at the buggy program with popular Python tools.
Figure~\ref{fig:error_tools} shows that the VSCode editor puts a red
squiggly line under variable \ycode{thetap} and PyLint explains that
it is undefined. Similarly, a green squiggly line warns that variable
\ycode{theta} is unused. If the code had used a string instead of a
variable, VSCode and PyLint would not have been able to detect and
explain these mistakes.

\begin{figure}
  \hspace*{-0.025\columnwidth}\includegraphics[width=1.05\columnwidth]{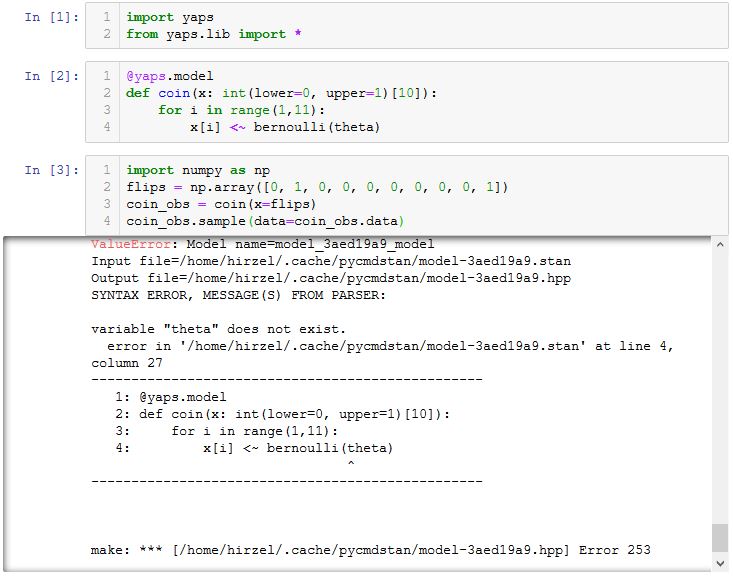}
  \caption{\label{fig:error_yaps}\yaps compiler source location mapping.}
  \vspace*{-2.5mm}
\end{figure}

For the second error-message experiment, we again injected an
undefined variable, then used \yaps and Stan to compile the code, see
Figure~\ref{fig:error_yaps}.  Stan finds the error in the compiled
code, and \yaps intercepts the error message from the Stan compiler,
mapping it back to the Python code. As is common in Jupyter notebooks,
the error message gets displayed as part of the cell output.

\section{Conclusion}\label{sec:concl}

This paper introduces \yaps, a new Python embedding of the Stan
probabilistic programming language that embraces Python syntax.
\yaps interfaces seamlessly with popular Py\-thon packages such as NumPy and
MatPlotLib, while at the same time remaining well-separated from them.
For testing purposes, we have done round-trip translations of over a
thousand models from Stan to \yaps and back. Future work includes
adding local variable type inference to make \yaps models leaner
without compromising their static typing guarantees; adding
meta-programming features so users can experiment with different
variants of their models; and possibly adding an interactive debugger.
\yaps is available at \url{https://github.com/ibm/yaps}.

\paragraph{Acknowledgements}
We thank Javier Burroni for insightful discussions and the Stan
team for their encouragement.

\bibliography{bibfile}


\begin{thebibliography}{18}


\ifx \showCODEN    \undefined \def \showCODEN     #1{\unskip}     \fi
\ifx \showDOI      \undefined \def \showDOI       #1{#1}\fi
\ifx \showISBNx    \undefined \def \showISBNx     #1{\unskip}     \fi
\ifx \showISBNxiii \undefined \def \showISBNxiii  #1{\unskip}     \fi
\ifx \showISSN     \undefined \def \showISSN      #1{\unskip}     \fi
\ifx \showLCCN     \undefined \def \showLCCN      #1{\unskip}     \fi
\ifx \shownote     \undefined \def \shownote      #1{#1}          \fi
\ifx \showarticletitle \undefined \def \showarticletitle #1{#1}   \fi
\ifx \showURL      \undefined \def \showURL       {\relax}        \fi
\providecommand\bibfield[2]{#2}
\providecommand\bibinfo[2]{#2}
\providecommand\natexlab[1]{#1}
\providecommand\showeprint[2][]{arXiv:#2}

\bibitem[\protect\citeauthoryear{Baudart, Hirzel, and Mandel}{Baudart
  et~al\mbox{.}}{2018}]%
        {baudart_hirzel_mandel_2018}
\bibfield{author}{\bibinfo{person}{Guillaume Baudart}, \bibinfo{person}{Martin
  Hirzel}, {and} \bibinfo{person}{Louis Mandel}.}
  \bibinfo{year}{2018}\natexlab{}.
\newblock \bibinfo{title}{Deep Probabilistic Programming Languages: A
  Qualitative Study}.
\newblock   (\bibinfo{year}{2018}).
\newblock
\showURL{%
\url{https://arxiv.org/abs/1804.06458}}


\bibitem[\protect\citeauthoryear{Breuleux and van Merri{\"e}nboer}{Breuleux and
  van Merri{\"e}nboer}{2017}]%
        {breuleux_vanmerrienboer_2017}
\bibfield{author}{\bibinfo{person}{Olivier Breuleux} {and}
  \bibinfo{person}{Bart van Merri{\"e}nboer}.} \bibinfo{year}{2017}\natexlab{}.
\newblock \showarticletitle{Automatic Differentiation in {Myia}}. In
  \bibinfo{booktitle}{{\em Autodiff Workshop}}.
\newblock
\showURL{%
\url{https://openreview.net/forum?id=S1hcluzAb}}


\bibitem[\protect\citeauthoryear{Burroni, Baudart, Mandel, Hirzel, and
  Shinnar}{Burroni et~al\mbox{.}}{2018}]%
        {burroni_et_al_2018}
\bibfield{author}{\bibinfo{person}{Javier Burroni}, \bibinfo{person}{Guillaume
  Baudart}, \bibinfo{person}{Louis Mandel}, \bibinfo{person}{Martin Hirzel},
  {and} \bibinfo{person}{Avraham Shinnar}.} \bibinfo{year}{2018}\natexlab{}.
\newblock \bibinfo{title}{Extending {Stan} for Deep Probabilistic Programming}.
\newblock   (\bibinfo{year}{2018}).
\newblock
\showURL{%
\url{https://arxiv.org/abs/1810.00873}}


\bibitem[\protect\citeauthoryear{Carpenter, Gelman, Hoffman, Lee, Goodrich,
  Betancourt, Brubaker, Guo, Li, and Riddell}{Carpenter et~al\mbox{.}}{2017}]%
        {carpenter_et_al_2017}
\bibfield{author}{\bibinfo{person}{Bob Carpenter}, \bibinfo{person}{Andrew
  Gelman}, \bibinfo{person}{Matt Hoffman}, \bibinfo{person}{Daniel Lee},
  \bibinfo{person}{Ben Goodrich}, \bibinfo{person}{Michael Betancourt},
  \bibinfo{person}{Michael~A. Brubaker}, \bibinfo{person}{Jiqiang Guo},
  \bibinfo{person}{Peter Li}, {and} \bibinfo{person}{Allen Riddell}.}
  \bibinfo{year}{2017}\natexlab{}.
\newblock \showarticletitle{Stan: A Probabilistic Programming Language}.
\newblock \bibinfo{journal}{{\em Journal of Statistical Software\/}}
  \bibinfo{volume}{76}, \bibinfo{number}{1} (\bibinfo{year}{2017}),
  \bibinfo{pages}{1--37}.
\newblock
\showURL{%
\url{https://www.jstatsoft.org/article/view/v076i01}}


\bibitem[\protect\citeauthoryear{Ghahramani}{Ghahramani}{2015}]%
        {ghahramani_2015}
\bibfield{author}{\bibinfo{person}{Zoubin Ghahramani}.}
  \bibinfo{year}{2015}\natexlab{}.
\newblock \showarticletitle{Probabilistic Machine Learning and Artificial
  Intelligence}.
\newblock \bibinfo{journal}{{\em Nature\/}} \bibinfo{volume}{521},
  \bibinfo{number}{7553} (\bibinfo{date}{May} \bibinfo{year}{2015}),
  \bibinfo{pages}{452--459}.
\newblock
\showURL{%
\url{https://www.nature.com/articles/nature14541}}


\bibitem[\protect\citeauthoryear{Gordon, Henzinger, Nori, and Rajamani}{Gordon
  et~al\mbox{.}}{2014}]%
        {gordon_et_al_2014}
\bibfield{author}{\bibinfo{person}{Andrew~D. Gordon},
  \bibinfo{person}{Thomas~A. Henzinger}, \bibinfo{person}{Aditya~V. Nori},
  {and} \bibinfo{person}{Sriram~K. Rajamani}.} \bibinfo{year}{2014}\natexlab{}.
\newblock \showarticletitle{Probabilistic Programming}. In
  \bibinfo{booktitle}{{\em ICSE track on Future of Software Engineering
  (FOSE)}}. \bibinfo{pages}{167--181}.
\newblock
\showURL{%
\url{https://doi.org/10.1145/2593882.2593900}}


\bibitem[\protect\citeauthoryear{Gorinova, Gordon, and Sutton}{Gorinova
  et~al\mbox{.}}{2018}]%
        {gorinova_gordon_sutton_2018}
\bibfield{author}{\bibinfo{person}{Maria~I. Gorinova},
  \bibinfo{person}{Andrew~D. Gordon}, {and} \bibinfo{person}{Charles Sutton}.}
  \bibinfo{year}{2018}\natexlab{}.
\newblock \showarticletitle{{SlicStan}: Improving Probabilistic Programming
  using Information Flow Analysis}. In \bibinfo{booktitle}{{\em Workshop on
  Probabilistic Programming Languages, Semantics, and Systems (PPS)}}.
\newblock
\showURL{%
\url{https://pps2018.soic.indiana.edu/files/2017/12/SlicStanPPS.pdf}}


\bibitem[\protect\citeauthoryear{Hudak}{Hudak}{1998}]%
        {hudak_1998}
\bibfield{author}{\bibinfo{person}{Paul Hudak}.}
  \bibinfo{year}{1998}\natexlab{}.
\newblock \showarticletitle{Modular Domain Specific Languages and Tools}. In
  \bibinfo{booktitle}{{\em International Conference on Software Reuse (ICSR)}}.
  \bibinfo{pages}{134--142}.
\newblock
\showURL{%
\url{https://doi.org/10.1109/ICSR.1998.685738}}


\bibitem[\protect\citeauthoryear{Hunter}{Hunter}{2007}]%
        {hunter_2007}
\bibfield{author}{\bibinfo{person}{John~D. Hunter}.}
  \bibinfo{year}{2007}\natexlab{}.
\newblock \showarticletitle{Matplotlib: A {2D} Graphics Environment}.
\newblock \bibinfo{journal}{{\em Computing In Science and Engineering
  (CISE)\/}} \bibinfo{volume}{9}, \bibinfo{number}{3} (\bibinfo{year}{2007}),
  \bibinfo{pages}{90--95}.
\newblock
\showURL{%
\url{https://doi.org/10.1109/MCSE.2007.55}}


\bibitem[\protect\citeauthoryear{P{\'e}rez}{P{\'e}rez}{2014}]%
        {perez_2014}
\bibfield{author}{\bibinfo{person}{Fernando P{\'e}rez}.}
  \bibinfo{year}{2014}\natexlab{}.
\newblock \bibinfo{title}{Project Jupyter}.
\newblock   (\bibinfo{year}{2014}).
\newblock
\showURL{%
\url{http://jupyter.org/}}
\newblock
\shownote{(Retrieved December 2018).}


\bibitem[\protect\citeauthoryear{Roesch, Lyubomirsky, Weber, Pollock, Kirisame,
  Chen, and Tatlock}{Roesch et~al\mbox{.}}{2018}]%
        {roesch_et_al_2018}
\bibfield{author}{\bibinfo{person}{Jared Roesch}, \bibinfo{person}{Steven
  Lyubomirsky}, \bibinfo{person}{Logan Weber}, \bibinfo{person}{Josh Pollock},
  \bibinfo{person}{Marisa Kirisame}, \bibinfo{person}{Tianqi Chen}, {and}
  \bibinfo{person}{Zachary Tatlock}.} \bibinfo{year}{2018}\natexlab{}.
\newblock \showarticletitle{Relay: A New {IR} for Machine Learning Frameworks}.
  In \bibinfo{booktitle}{{\em Workshop on Machine Learning and Programming
  Languages (MAPL)}}. \bibinfo{pages}{58--68}.
\newblock
\showURL{%
\url{http://doi.acm.org/10.1145/3211346.3211348}}


\bibitem[\protect\citeauthoryear{Rompf and Odersky}{Rompf and Odersky}{2012}]%
        {rompf_odersky_2012}
\bibfield{author}{\bibinfo{person}{Tiark Rompf} {and} \bibinfo{person}{Martin
  Odersky}.} \bibinfo{year}{2012}\natexlab{}.
\newblock \showarticletitle{Lightweight Modular Staging: A Pragmatic Approach
  to Runtime Code Generation and Compiled {DSLs}}.
\newblock \bibinfo{journal}{{\em Communications of the ACM (CACM)\/}}
  \bibinfo{volume}{55} (\bibinfo{year}{2012}), \bibinfo{pages}{121--130}.
\newblock
Issue 6.
\showURL{%
\url{https://doi.org/10.1145/2184319.2184345}}


\bibitem[\protect\citeauthoryear{Salvatier, Wiecki, and Fonnesbeck}{Salvatier
  et~al\mbox{.}}{2015}]%
        {salvatier_wiecki_fonnesbeck_2015}
\bibfield{author}{\bibinfo{person}{John Salvatier}, \bibinfo{person}{Thomas~V.
  Wiecki}, {and} \bibinfo{person}{Christopher Fonnesbeck}.}
  \bibinfo{year}{2015}\natexlab{}.
\newblock \bibinfo{title}{Probabilistic Programming in {Python} Using {PyMC3}}.
\newblock   (\bibinfo{year}{2015}).
\newblock
\showURL{%
\url{https://arxiv.org/abs/1507.08050}}


\bibitem[\protect\citeauthoryear{Tran, Hoffman, Saurous, Brevdo, Murphy, and
  Blei}{Tran et~al\mbox{.}}{2017}]%
        {tran_et_al_2017}
\bibfield{author}{\bibinfo{person}{Dustin Tran}, \bibinfo{person}{Matthew~D.
  Hoffman}, \bibinfo{person}{Rif~A. Saurous}, \bibinfo{person}{Eugene Brevdo},
  \bibinfo{person}{Kevin Murphy}, {and} \bibinfo{person}{David~M. Blei}.}
  \bibinfo{year}{2017}\natexlab{}.
\newblock \showarticletitle{Deep Probabilistic Programming}. In
  \bibinfo{booktitle}{{\em International Conference on Learning Representations
  (ICLR)}}.
\newblock
\showURL{%
\url{https://arxiv.org/abs/1701.03757}}


\bibitem[\protect\citeauthoryear{Uber}{Uber}{2017}]%
        {uber_2017}
\bibfield{author}{\bibinfo{person}{Uber}.} \bibinfo{year}{2017}\natexlab{}.
\newblock \bibinfo{title}{Pyro}.
\newblock   (\bibinfo{year}{2017}).
\newblock
\showURL{%
\url{http://pyro.ai/}}
\newblock
\shownote{(Retrieved December 2018).}


\bibitem[\protect\citeauthoryear{{van der Walt}, Colbert, and Varoquaux}{{van
  der Walt} et~al\mbox{.}}{2011}]%
        {vanderwalt_colbert_varoquaux_2011}
\bibfield{author}{\bibinfo{person}{St{\'{e}}fan {van der Walt}},
  \bibinfo{person}{S.~Chris Colbert}, {and} \bibinfo{person}{Ga{\"{e}}l
  Varoquaux}.} \bibinfo{year}{2011}\natexlab{}.
\newblock \showarticletitle{The {NumPy} Array: A Structure for Efficient
  Numerical Computation}.
\newblock \bibinfo{journal}{{\em Computing in Science and Engineering
  (CISE)\/}} \bibinfo{volume}{13}, \bibinfo{number}{2} (\bibinfo{year}{2011}),
  \bibinfo{pages}{22--30}.
\newblock
\showURL{%
\url{https://doi.org/10.1109/MCSE.2011.37}}


\bibitem[\protect\citeauthoryear{Wiltschko, van Merri{\"e}nboer, and
  Moldovan}{Wiltschko et~al\mbox{.}}{2017}]%
        {wiltschko_vanmerrienboer_moldovan_2017}
\bibfield{author}{\bibinfo{person}{Alex Wiltschko}, \bibinfo{person}{Bart van
  Merri{\"e}nboer}, {and} \bibinfo{person}{Dan Moldovan}.}
  \bibinfo{year}{2017}\natexlab{}.
\newblock \bibinfo{title}{Tangent: Source-to-Source Debuggable Derivatives in
  Pure {Python}}.
\newblock   (\bibinfo{year}{2017}).
\newblock
\showURL{%
\url{https://github.com/google/tangent}}
\newblock
\shownote{(Retrieved December 2018).}


\bibitem[\protect\citeauthoryear{Woodman}{Woodman}{2017}]%
        {woodman_2017}
\bibfield{author}{\bibinfo{person}{Marmaduke Woodman}.}
  \bibinfo{year}{2017}\natexlab{}.
\newblock \bibinfo{title}{{PyCmdStan}: {Python} Interface to {CmdStan}}.
\newblock   (\bibinfo{year}{2017}).
\newblock
\showURL{%
\url{https://gitlab.thevirtualbrain.org/tvb/pycmdstan}}
\newblock
\shownote{(Retrieved December 2018).}


\end{thebibliography}

\end{document}